\documentclass[twocolumn,showpacs,preprintnumbers,eqsecnum,amsmath,amssymb]{revtex4}

\usepackage{graphicx}%
\usepackage{dcolumn}%
\usepackage{bm}%
\usepackage{epsf}%

\def\tV{{\tilde V}}
\def\tU{{\tilde U}}

\def\cF{{\cal F}}
\def\cG{{\cal G}}
\def\cH{{\cal H}}
\def\cK{{\cal K}}
\def\cL{{\cal L}}
\def\cOi{{\cal O}^\infty}
\def\cO{{\cal O}}

\def\oP{{\overline P}}

\def\oW{{{\overline W}_\eta}}
\def\dR{{\dot R}}
\def\dW{{{\dot W}_\eta}}
\def\Rp{{R'}}
\def\Wp{{{W_\eta}'}}
\def\zr{{\mathfrak r}}
\def\zp{{\mathfrak p}}
\def\Gin{{{\Gamma_\eta}_{\rm in}}}
\def\Gout{{{\Gamma_\eta}_{\rm out}}}

\begin{document}
\preprint{UATP-01/06}
\title{Toward a Quantization of Null Dust Collapse\\}
\author{Cenalo Vaz${}^{a}$ and Louis Witten$^{b}$ and T.P. Singh${}^{c}$ \\}

\affiliation{$^{a}$Faculdade de Ci\^encias e Tecnologia, \\
Universidade do Algarve, Faro, Portugal.\\
{\rm Email address: {\tt cvaz@ualg.pt}}}

\affiliation{$^{b}$Department of Physics,\\
University of Cincinnati, Cincinnati, OH 45221-0011, USA.\\
{\rm Email address: {\tt witten@physics.uc.edu}}}

\affiliation{$^{c}$Tata Institute of Fundamental Research,\\
Homi Bhabha Road, Mumbai 400 005, India.\\
{\rm Email address: {\tt tpsingh@nagaum.tifr.res.in}}}

\begin{abstract}
Spherically symmetric, null dust clouds, like their time-like counterparts, may collapse
classically into black holes or naked singularities depending on their initial conditions.
We consider the Hamiltonian dynamics of the collapse of an arbitrary distribution of null
dust, expressed in terms of the physical radius, $R$, the null coordinates, $V$ for a
collapsing cloud or $U$ for an expanding cloud, the mass function, $m$, of the null matter,
and their conjugate momenta. This description is obtained from the ADM description by a
Kucha\v r-type canonical transformation. The constraints are linear in the canonical momenta
and Dirac's constraint quantization program is implemented. Explicit solutions to the
constraints are obtained for both expanding and contracting null dust clouds with arbitrary
mass functions.

\end{abstract}

\maketitle

\section{\label{sec:intro}Introduction}
Spherically symmetric dust clouds,  depending on their initial matter and velocity distributions,
will collapse in classical general relativity to form either black holes or naked singularities.
Black holes are better understood than naked singularities. They are generally expected to evaporate
via their associated Hawking radiation \cite{hawk1}, although no agreement has yet been achieved
regarding the end state of collapse, {\it i.e.,} whether a remnant survives or whether all the matter
contained in the original cloud is thermally radiated away. If a portion of the collapsing matter
does manage to form a stable black hole, it is expected that the total mass of the remnant will be
quantized. On the other hand, if all the matter is radiated away before a stable end state can form
then one must explain what happens to the information that was contained in the initial matter
distribution. The formation of black holes therefore presents a number of deep puzzles and various
approaches to quantum gravity are being employed to address these at the present time
\cite{bek1,av,st,cqg}. On the contrary, naked singularities have received comparatively little attention.
Yet, the formation of naked singularities (singularities that are visible either locally or asymptotically)
is much more difficult to understand and for an entirely different reason: their existence implies the
absence of a well defined Cauchy problem to the future of some light-like surface (the Cauchy horizon),
therefore any attempt to describe the system to the future of this surface fails for lack of initial
conditions. It seems that space-time must be terminated at the Cauchy horizon. In order to avoid the
associated problems, Penrose proposed a Cosmic Censor \cite{Rp1}, whose function is essentially to
ensure that naked singularities never form. The mechanism by which the Cosmic Censor operates, however,
is still shrouded in mystery. The Censor is most likely not classical because most models of classical
collapse lead to the formation of both black holes and naked singularities in different domains of
the initial phase space \cite{Pj1}. In fact very little is currently understood about the final stages
of a collapse that leads to the formation of a classical naked singularity.

There are indications from the semi-classical treatment of naked singularities, in which the
gravitational degrees of freedom are considered to be classical, that Penrose's Cosmic Censor may, in
fact, be the quantum theory itself \cite{cvetc}. However, at the very final stages of collapse it is
not possible to treat the gravitational degrees of freedom classically and a full blown quantum theory
of the gravitational field becomes necessary to establish this possibility \cite{haretc} firmly.
Singularities in general relativity signal a breakdown of the classical theory, a regime in which
the classical equations are meaningless.  Cosmic Censorship probably points to the need for
quantum gravity in the same way as, more than eighty years ago, the electrodynamic instability of
atoms pointed to the need for quantum mechanics. A good question is just how complete a theory of
quantum gravity is required to begin addressing such issues as the Cosmic Censor. We take the attitude
that, from past experience, it is not unreasonable to expect many of the key effects of quantum gravity
to be understood from a more na\"\i ve quantization of the gravitational field which, while it may be
incomplete, incorporates the essential features of quantum mechanics.

This is what we propose to do in this paper. Our objective is to consider the midi-superspace
quantization of a spherically symmetric cloud of null matter specified by an arbitrary mass
distribution and collapsing in its own gravitational field. The model we are concerned with therefore
is a solution of Einstein's equations with pressureless, null dust \cite{vaidya} described by the stress
energy tensor $T_{\mu\nu} = \epsilon(x) U_\mu U_\nu$, where $\epsilon(x)$ is the energy density of
the cloud and $U^2 = 0$. When the cloud is contracting, the solution is characterized by an arbitrary
function, $m(V)$, of the advanced null coordinate, $V \in (-\infty,+\infty)$. The mass function is
generally taken to be vanishing for $V < V_o$ and constant, $M$, when $V > V_1$. The space-time is
described by the metric
\begin{equation}
ds^2 = \left(1-\frac{2m(V)}{R}\right) dV^2 - 2dRdV - R^2 d\Omega^2,
\label{vaidyain}
\end{equation}
where $R \in [0,+\infty)$ is the area radius. The region $V > V_1$ is a part of the Schwarzschild
space-time. In this region the metric may be written in terms of the Eddington-Finkelstein coordinates,
$\tU$ and $\tV$ as
\begin{equation}
ds^2 = \left(1-\frac{2M}{R}\right) d\tU d\tV - R^2 d\Omega^2,
\label{vaidyaout}
\end{equation}
where $\tU=\tV-2R^*$ and $R^*$ is the tortoise coordinate. The region $V<V_o$ is a part of Minkowski
space-time, with metric
\begin{equation}
ds^2 = du dv - R^2 d\Omega^2,
\end{equation}
where $u$ and $v$ are the ordinary retarded and advanced times, respectively $T \mp R$.

In the time reversed situation the null dust cloud is expanding instead of contracting and the
solution is written in an analogous fashion,
\begin{equation}
ds^2 = \left(1-\frac{2m(U)}{R}\right) dU^2 + 2dRdU - R^2 d\Omega^2
\end{equation}
in terms of a {\it retarded} null coordinate, $U$. Again, the mass function is generally taken to
vanish when $U > U_1$, having some constant value, $M$ before some earlier retarded time, $U < U_o$.
The region $U > U_1$ is then a part of Minkowski space-time, while the region $U < U_o$ is a part
of the Schwarzschild space-time.

Depending on the distribution $m(V)$ of matter in the null cloud, either black holes or
naked singularities may develop as the classical final state of the collapse. For example, in the self
similar model in which the mass is a linear function of the advanced null coordinate, $m(V) = \lambda V$,
one finds that both outcomes described by the Penrose diagrams in figure 1 are possible, depending on
whether $\lambda > 1/16$ (black hole) or $\lambda \leq 1/16$ (naked singularity) \cite{nsing}.
\begin{figure}
\epsfysize=5cm \epsfbox{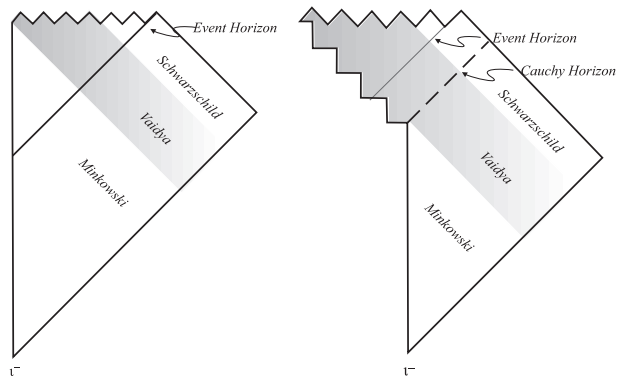}
\caption{Black hole (left) or naked singularity (right) formation in null dust collapse}
\end{figure}
When the collapse evolves toward a naked singularity, spatial hypersurfaces in the future of the
initial singularity cross the Cauchy horizon and collide with the central singularity but, because
no sensible boundary conditions can be specified on a singularity, the evolution in the future of the
initial singularity is arbitrary. The Cosmic Censor \cite{Rp1} should come into play {\it before}
the Cauchy horizon has a chance to form. It is of interest, therefore, to understand how the system
behaves close to, but in the past of, the putative Cauchy horizon, where spatial hypersurfaces are
well defined and the quantum evolution of the system may be studied.

Null shells classically collapse to form covered singularities, therefore, in order to examine such
issues as the Cosmic Censor, mass distributions other than those representing shells should be considered.
These require a quantization of a genuine field theory. The present paper is intended as a first step
in this direction. The solution metric is written in Eddington-Finkelstein coordinates and the Kucha\v r
transformation from ADM variables to Kucha\v r variables is established. The mass function is explicitly
related to $m(V)$ and $m(U)$ of the metrics in (\ref{vaidyain}), respectively (\ref {vaidyaout}).

The quantization program employed in this paper involves a gauge fixing and it is known that quantum
theories resulting from different gauge fixings are not necessarily equivalent. It is, nevertheless,
the best approach to the quantization of realistic and pressing problems such as gravitational
collapse at this time. Our choice of configuration space coordinates presents several advantages:
their physical meaning is transparent, we are able to give explicit transformations between the
original ADM phase space and the new, and the constraints expressed in terms of the new phase space
variables are linear. The last enables us to obtain solutions for null dust clouds of arbitrary mass
distributions. The second implies that operators representing observables, if known in one system
are easily constructed in the other, and the first makes our solutions easily interpretable in physical
terms.

In section \ref{sec:adm} we summarize the canonical formulation of the action in ADM variables. The
null dust action appropriate to the models being considered is  also analyzed in this section. In
section \ref{sec:cantr} we explicitly perform a transformation of the phase-space to Kucha\v r variables
\cite{ku1}. In section \ref{sec:quant} we apply Dirac's quantization program to these models. Solutions
to the constraints for collapsing and expanding clouds are presented and the matter distribution
representing a single shell is examined as a special case. However, the phase-space admits non-trivial
boundaries and this generally leads to complications in the quantization procedure. Our solutions are
valid subject to the condition that a suitable measure can be found so that the replacement
$P_X \rightarrow -i\delta/\delta X(r)$ leads to self-adjoint operators. We conclude in section \ref{sec:disc}
with a discussion of the strengths and weaknesses of our approach, the issues that remain to be
resolved (such as the question of observables and of the measure on the Hilbert space) as well as some
suggestions for future directions.

\section{\label{sec:adm}Canonical Formulation in ADM Variables}

Consider the line element, $d\sigma$ on a spherically symmetric three dimensional Riemann surface,
$\Sigma$. It is completely characterized by two functions, $L(r)$ and $R(r)$ of the radial label
coordinate
\begin{equation}
ds_{(3)}^2 = L^2(r) dr^2 + R^2(r) d\Omega^2,
\label{hyper}
\end{equation}
where $\Omega$ is the solid angle. The angular coordinates play no role and will be integrated over.
We take both $L(r)$ and $R(r)$ to be positive definite except, possibly, at the center. $R(r)$
represents the physical radius of a shell labeled by $r$ on the surface. It behaves as a scalar
under transformations of $r$, whereas $L(r)$ behaves as a scalar density. The corresponding four
dimensional line element may be written in terms of two additional functions, the lapse, $N(t,r)$,
and the shift, $N^r(t,r)$, as
\begin{equation}
ds^2~ =~ N^2 dt^2 - L^2(dr + N^r dt)^2 - R^2 d\Omega^2.
\label{gsmet}
\end{equation}
In this spherically symmetric space-time, we will consider the Einstein-Dust system
described by the action
\begin{eqnarray}
S = & - & \frac{1}{16\pi} \int d^4 x \sqrt{-g}~ {\cal R}\cr
    & - & \frac{1}{8\pi} \int d^4 x \sqrt{-g}~ \epsilon(x)g^{\alpha\beta}
        U_\alpha U_\beta
\label{action}
\end{eqnarray}
where ${\cal R}$ is the scalar curvature. As is well known, the gravitational part of this
action can be cast into the form
\begin{eqnarray}
S^g =  \int  &dt& \int_0^\infty dr \left [ P_L {\dot L} + P_R {\dot R} - N H^g
- N^r H^g_r \right]\cr\cr
& + &~ S^g_{\partial \Sigma}
\end{eqnarray}
with the momenta conjugate to $L$ and $R$ respectively given by
\begin{eqnarray}
P_L~~ &=&~~ \frac{R}{N} \left[-{\dot R} + N^r R'\right]\cr
P_R~~ &=&~~ \frac{1}{N} \left[-L{\dot R} - {\dot L} R + (N^r LR)'\right]
\label{gravmom}
\end{eqnarray}
and where the overdot and the prime refer respectively to partial derivatives with respect
to the label time, $t$, and coordinate, $r$. The lapse, shift and phase-space variables are
required to be continuous functions of the label coordinates. The boundary action,
$S^g_{\partial \Sigma}$, is required to cancel unwanted boundary terms in the hypersurface
action, ensuring that the hypersurface evolution is not frozen on the frontiers. It is determined
after fall-off conditions appropriate to the models under consideration are specified. The
super-Hamiltonian and super-momentum constraints are given by
\begin{eqnarray}
\cH^g = &-& \left[\frac{P_L P_R}{R} - \frac{LP_L^2}{2R^2}\right]\cr
&+& \left[ - \frac{L}{2} - \frac{{R'}^2}{2L} + \left(\frac{RR'}{L}\right)'\right]\cr\cr
\cH^g_r = &+&R' P_R - L P_L'.
\label{const}
\end{eqnarray}
We will assume that the matter distribution is such that at infinity Kucha\v r's fall-off
conditions \cite{ku1} are suitable and we will adopt them here. These conditions would
be applicable, for example, in models in which the collapsing metric asymptotically approaches
or is smoothly matched to an exterior Schwarzschild background at some boundary. They
read
\begin{eqnarray}
L(t,r)~~ &=& ~~ 1~ +~ M_+(t) r^{-1}~ +~ \cOi(r^{-1-\epsilon})\cr
R(t,r)~~ &=& ~~ r~ +~ \cOi(r^{-\epsilon})\cr
P_L(t,r)~~ &=& ~~ \cOi(r^{-\epsilon})\cr
P_R(t,r)~~ &=&~~ \cOi(r^{-1-\epsilon})\cr
N(t,r)~~ &=& ~~ N_+(t)~ +~ \cOi(r^{-\epsilon})\cr
N^r(t,r)~~ &=&~~ \cOi(r^{-\epsilon})
\label{foinf}
\end{eqnarray}
and imply that the asymptotic regions are flat with the spatial hypersurfaces asymptotic to
surfaces of constant Minkowski time. Again, as $r \rightarrow 0$ we require that \cite{Louko}
\begin{eqnarray}
L(t,r)~~ &=& ~~ L_0(t)~ +~ \cO(r^2)\cr
R(t,r)~~ &=& ~~ R_1(t) r~ +~ \cO(r^3)\cr
P_L(t,r)~~ &=& ~~ P_{L_2}(t) r^2~ +~ \cO(r^4)\cr
P_R(t,r)~~ &=&~~ P_{R_1}(t)r~ +~ \cO(r^3)\cr
N(t,r)~~ &=& ~~ N_0(t)~ +~ \cO(r^2)\cr
N^r(t,r)~~ &=&~~ N^r_1(t) r~ +~ \cO(r^3)
\label{fo0}
\end{eqnarray}
With these conditions, it is easy to see that the appropriate choice of surface action
involves only the contribution,
\begin{equation}
S^g_{\partial \Sigma} = - \int_{\partial \Sigma_\infty} dt N_+(t) M_+(t)
\end{equation}
at the boundary at infinity.

Let us now consider the null dust action in (\ref{action}). We note first that the energy
density, $\epsilon(x)$, plays the role of a Lagrange multiplier enforcing null dust, {\it
i.e.,} $U^2 = 0$, and variation w.r.t. $g_{\alpha\beta}$ yields the standard dust stress tensor,
$T_{\alpha\beta} = \epsilon(x) U^\alpha U^\beta$. The canonical form of the null dust action
in various forms has been studied by Kucha\v r and Bi\v c\'ak \cite{kuBi}. In particular, for
the action in the form given in (\ref{action}) one may expand $U_\alpha$ as a Pfaff form of six
scalar fields, the three co-moving coordinates of the null dust particles, $Z^k$, and three
scalars (velocities), $w_k$,
\begin{equation}
U_\alpha = w_k Z^k_{,\alpha}.
\end{equation}
This representation is redundant because, by Pfaff's theorem, only four scalars are required
to describe an arbitrary covector in a four dimensional space. Suppose we require one of the
scalars, say $w_3$ to be unity and drop the index from the associated co-moving coordinate,
$Z^3 := Z$, then
\begin{equation}
U_\alpha = Z_{,\alpha} + w_k Z^k_{,\alpha},~~~~ k \in \{1,2\}.
\label{ualpha}
\end{equation}
Consider the independent variations,
\begin{eqnarray}
0 &=& \frac{\delta S}{\delta \epsilon} = -\sqrt{-g} g^{\alpha\beta} U_\alpha U_\beta\cr\cr
0 &=& \frac{\delta S}{\delta Z} = (\sqrt{-g}\epsilon(x) U^\alpha)_{,\alpha} = \nabla_\alpha
(\epsilon(x) U^\alpha)\cr\cr
0 &=& \frac{\delta S}{\delta w_k} = -\sqrt{-g}\epsilon(x) g^{\alpha\beta} Z^k_{,\alpha} U_\beta
\cr\cr
0 &=& \frac{\delta S}{\delta Z^k} = (\sqrt{-g}\epsilon(x)w_k U^\alpha)_{,\alpha}\cr\cr &=&
\nabla_\alpha (\epsilon(x) w_k U^\alpha)\cr\cr
T_{\alpha\beta} &=& \frac{2}{\sqrt{-g}}\frac{\delta S}{\delta g_{\alpha\beta}} = \epsilon(x)
U_\alpha U_\beta
\label{dustvars}
\end{eqnarray}
The conservation of the stress energy tensor in the last equation implies that
\begin{equation}
\epsilon(x) U^\alpha \nabla_\alpha U^\beta + U^\beta \nabla_\alpha (\epsilon(x) U^\alpha) = 0,
\end{equation}
which says that the particles follow geodesic curves. Using the second equation above we find
$U^\alpha \nabla_\alpha U^\beta = 0$, implying affine parameterization. The third equation says
that $\cL_U Z^k = Z^k_{,\alpha}U^\alpha = 0$, {\it i.e.,} all of the $Z^k$ are constant along
flow lines and none of them are time-like. And finally, multiplying the third equation by $w_k$ we
find
\begin{equation}
Z_{,\alpha}U^\alpha = 0,
\end{equation}
saying that $Z_{,\alpha}$ may be space-like or null. If the twist, $U_{[\alpha;\beta]}$, also
vanishes, then $Z_{,\alpha}$ is null, which would imply that $w_k Z^k_{,\alpha} = 0$, or
$w_k = 0~ \forall~ k \in \{1,2\}$, because the $Z^k_{,\alpha}$ are taken to form a (linearly
independent) cobasis.

Substituting the decomposition (\ref{ualpha}) into the dust action in (\ref{action}), using
(\ref{gsmet}) and integrating over the angular coordinates the action may be put in the form
\begin{equation}
S^d = \int dt \int_0^\infty dr \left[P_Z {\dot Z}  + P_k {\dot Z}^k - N \cH^d - N^r \cH^d_r\right],
\end{equation}
where the momenta conjugate to $\{Z,Z^k\}$ are, respectively,
\begin{eqnarray}
P_Z &=& \frac{L R^2}{N}\epsilon(r) [({\dot Z} + w_k {\dot Z}^k) - N^r (Z' + w_k {Z^k}')]\cr\cr
P_k &=& w_k P_Z,
\end{eqnarray}
and the constraints, $\cH^d$ and $\cH^d_r$ are
\begin{eqnarray}
\cH^d &=& \left[\frac{P_Z^2}{2LR^2 \epsilon} + \frac{\epsilon R^2 (Z' + w_k {Z^k}')^2}{2L}\right]
\cr\cr
\cH^d_r &=& P_Z (Z' + w_k {Z^k}').
\end{eqnarray}
Setting $\delta \cL / \delta \epsilon = 0$ gives the final form of the dust Hamiltonian and
momentum constraints
\begin{eqnarray}
\cH^d &=& \pm \frac{P_Z(Z' + w_k {Z^k}')}{L}\cr\cr
\cH^d_r &=& P_Z (Z' + w_k {Z^k}'),
\end{eqnarray}
where the positive (respectively negative) signs in the dust Hamiltonian density represent incoming
(respectively outgoing) dust. In the spherically symmetric collapse we are considering, we take
$w_k = 0 = P_k$. Thus we have arrived at the canonical form of our theory, which we will write as
\begin{eqnarray}
S = &+& \int dt \int_0^r dr [{\dot Z} P_Z + {\dot L} P_L + \cr\cr
&+& \dR P_R - N \cH - N^r \cH_r]\cr\cr
\cH = &-& \left[\frac{P_L P_R}{R} - \frac{LP_L^2}{2R^2}\right]\cr
&+& \left[ - \frac{L}{2} - \frac{{R'}^2}{2L} + \left(\frac{RR'}{L}\right)'\right] -
\eta \frac{P_Z Z'}{L}\cr\cr
\cH^g_r = && Z'P_Z  + R' P_R - L P_L'\cr\cr
S_{\partial\Sigma} = &-& \int_{\partial\Sigma_\infty} N_+(t)M_+(t),
\label{system}
\end{eqnarray}
where $\eta = {\rm sgn}(Z')$. In the following section we will show that the co-moving coordinate $Z$
may be identified with the null coordinates according to $Z=-U$ for an expanding solution and $Z=-V$
for a collapsing one. $P_Z \leq 0$ and the dust Hamiltonian density is chosen to be always non-negative.
When $P_Z$ is non-vanishing the phase-space is made up of two disconnected sectors, labeled by $\eta$.
An initial data set with $P_Z = 0$ cannot evolve into a set with $P_Z \neq 0$ and we will assume from
now on that $P_Z \neq 0$.

\section{\label{sec:cantr}Canonical Transformation}

The description of contracting and expanding clouds is seen to be related by time
reversal. The two descriptions may be formally unified in the following way. Introduce a null
coordinate $W_\eta$, which can be the ``advanced'' time or the ``retarded'' time, satisfying only the
requirement that $W_\eta$ increases toward the future. If $W_\eta' < 0$ (primes denote differentiation
w.r.t. the ADM label coordinate $r$) then it represents the retarded coordinate, $U$, and if, on the
contrary, $W_\eta' > 0$, it represents the advanced coordinate, $V$. Let us write both solutions in
terms of a parameter $\eta$ that represents the behavior of the matter (whether it is expanding or
collapsing)
\begin{equation}
ds^2 = \left(1 - \frac{2m}{R}\right) d{W_\eta}^2 + 2 \eta d{W_\eta} dR - R^2 d\Omega^2.
\label{sptime}
\end{equation}
The metric (\ref{sptime}) is appropriate for either expansion or contraction of the dust cloud depending
on whether $\eta = -{\rm sgn}({W_\eta}')$ is $+1$ or $-1$. $\eta$ is the same as appears in (\ref{system}),
as we argue below. The null coordinate, $\oW$, whose spatial direction is opposite to ${W_\eta}$ is
obtained by integrating
\begin{equation}
d\oW = \sigma({W_\eta},R)\left[d{W_\eta} + 2\eta \frac{dR}{\cF}\right],
\end{equation}
where $\sigma({W_\eta},R)$ is an integrating factor. This coordinate must also be always increasing
toward the future.

The hypersurfaces (\ref{hyper}) from which (\ref{gsmet}) is constructed must be embedded
in the space-time described by the metric (\ref{sptime}). Substituting the foliation ${W_\eta}(t,r)$
and $R(t,r)$ in (\ref{sptime}) gives the density $L(t,r)$ and the lapse and shift functions, $N(t,r)$
and $N^r(t,r)$, as
\begin{eqnarray}
\cF \dW^2 + 2\eta \dW \dR &=& N^2 - L^2 {N^r}^2\cr
-\cF \Wp^2 - 2\eta \Wp\Rp &=& L^2\cr
-\eta(\dW \Rp + \Wp \dR) - \cF \dW \Wp &=& {N^r} L^2,
\label{trans1}
\end{eqnarray}
where we have set $\cF = 1 - 2m/R$. These relations can be used to determine,
\begin{eqnarray}
N^r &=& \frac{\cF{\dot {W_\eta}} {W_\eta}' + \eta({\dot {W_\eta}} R' + {W_\eta}'{\dot R})}{\cF{
{W_\eta}'}^2 + 2\eta {W_\eta}' R'}\cr
N &=& \frac{{\dot {W_\eta}} R' - {{W_\eta}'}{\dot R}}{L},
\label{trans2}
\end{eqnarray}
where we have chosen the sign of the square root so that $N$ is positive. Inserting these expressions
for $N$ and $N^r$ in the expression (\ref{gravmom}) for $P_L$, we find
\begin{equation}
\frac{LP_L}{R} = -\eta R' - \cF {W_\eta}',
\label{wprime}
\end{equation}
which, when substituted into the expression for $L^2$ in (\ref{trans1}), gives
\begin{equation}
\cF = \frac{{R'}^2}{L^2} - \frac{P_L^2}{R^2},
\end{equation}
or, equivalently, the mass function in terms of the canonical data
\begin{equation}
m = \frac{R}{2} \left[1 - \frac{{R'}^2}{L^2} + \frac{P_L^2}{R^2}\right].
\label{mass}
\end{equation}
By directly taking Poisson brackets, the momentum conjugate to $m$ can now be shown to be simply
\begin{equation}
P_m = \frac{LP_L}{R\cF}.
\end{equation}
Kucha\v r\cite{ku1} proposed that $(R,m,\oP_R,P_m)$ should form a canonical chart whose coordinates
are spatial scalars, whose momenta are scalar densities and which is such that $H_r(r)$
generates Diff {\bf R}. This means that
\begin{equation}
\cH^g_r =  R'P_R - L P_L' = R'\oP_R + m'P_m \approx 0.
\end{equation}
Substituting the expressions derived for $m$ and $P_m$ into the above constraint one arrives at
\begin{equation}
\oP_R = P_R - \frac{LP_L}{2R} - \frac{LP_L}{2R\cF} - \frac{\Delta}{RL^2 \cF},
\label{op}
\end{equation}
where $\Delta = (RR')(LP_L)' - (RR')'(LP_L)$. One can then show that the transformation,
\begin{equation}
(R,L,P_R,P_L) \rightarrow (R,m,\oP_R,P_m),
\end{equation}
is canonical, and generated by
\begin{equation}
\cG = \int_0^\infty dr \left[LP_L - \frac{1}{2}{RR'} \ln |\frac{RR'+LP_L}{RR'-LP_L}|
\right]
\label{genfn}
\end{equation}
By computing the difference between the old and the new Liouville forms and using the fall off
conditions in (\ref{foinf}) and (\ref{fo0}), one can show that the transformation has introduced
no fresh boundary terms.

There are (infinite) boundary terms at the horizon, when $\cF = 0$. It can be shown, however, that
the contribution from the interior and the exterior cancel each other. There will also be
contributions at the boundary between the interior of the star and its exterior or more generally
at any frontier between two regions described by mass functions with different derivatives. Again,
if the mass function is continuous across the boundary and regions are consistently matched by
equating both the first and second fundamental forms, then the contribution from one side will
cancel the contribution from the other.

Before rearranging the action, we will consider the coordinates $\{Z,Z^k\}$ for the
collapsing Vaidya null congruence. Returning to the metric in (\ref{sptime}) with $\eta = -1$
we find that the incoming null congruence is given by $V={\rm const.}$, $\theta={\rm const.}$ and
$\phi={\rm const.}$ The coordinates $Z=-V$, $Z^1 = \theta$ and $Z^2 = \phi$ are co-moving. Let
us form the basis
\begin{eqnarray}
Z_{,\mu} &=& (-1,0,0,0)\cr
Z^1_{,\mu} &=& (0,0,1,0)\cr
Z^2_{,\mu} &=& (0,0,0,1)
\end{eqnarray}
It is easily shown that $\xi = -R$ is an affine parameter and the covariant components of
the velocity $dx^\mu/d\xi$ are $U_\mu = (-1,0,0,0)$, whose decomposition in the co-basis
$Z^k_{,\alpha}$ yields $W_1 = 0 = W_2$. Similarly treating the outgoing null congruence shows
that the affine parameter is $\xi = +R$ and that $Z$ is to be identified with $-U$. Both cases
may be treated simultaneously by letting $Z = -{W_\eta}$, $P_Z=-P_{W_\eta}$ and $\xi=\eta R$.
This identification shows that the $\eta$ used in the section is identical to that used in the
previous section.

Note also that taking the spatial derivative of $m$ in (\ref{mass}), yields
\begin{equation}
m' = - \frac{R'}{L} \cH^g - \frac{P_L}{RL} \cH^g_r.
\label{mprime}
\end{equation}
This may be used to write the action in (\ref{action}) as
\begin{widetext}
\begin{equation}
S = \int dt \int_0^\infty dr \left[P_{W_\eta} {\dot {W_\eta}} + \oP_R {\dot R} + P_m {\dot m}
- N \cH - N^r \cH_r\right] + S_{\partial\Sigma},
\label{system1}
\end{equation}
\end{widetext}
with
\begin{eqnarray}
\cH &=& - \left[\frac{m' \cF^{-1} R' + \cF P_m \oP_R}{L} \right] - \eta \frac{P_{W_\eta} {W_\eta}'}
{L}\cr\cr
\cH_r &=& R'\oP_R + m' P_m + {W_\eta}' P_{W_\eta}\cr\cr
S_{\partial\Sigma} &=& -\int_{\partial\Sigma_\infty} N_+(t)M_+(t).
\label{const1}
\end{eqnarray}
The surface term contains the mass at spatial infinity and may be re-expressed in a more convenient form. Use
the fall-off conditions (\ref{foinf}) at infinity and the expression for $N$ in (\ref{trans2}) to
write $N_+ = {\dot {W_\eta}}_+$ and
\begin{equation}
S_{\partial \Sigma} = -\int_{\partial \Sigma_\infty} {\dot {W_\eta}}_+ (t) M_+(t).
\end{equation}
Then define ${\Gamma_\eta}(r)$ by
\begin{equation}
m(r) = M_+ - \int_r^\infty dr' {\Gamma_\eta}(r')
\label{mredef}
\end{equation}
{\it i.e.,} $m'(r) = {\Gamma_\eta}(r)$ and $M_+$ is clearly the mass at infinity. The one form
\begin{equation}
\Omega = \int_0^\infty dr P_m(r) \delta m(r) - M_+ \delta {W_\eta}_+
\end{equation}
can now be written as
\begin{eqnarray}
\Omega &=& \left({W_\eta}_+ + \int_0^\infty dr P_m(r)\right)\delta M_+\cr\cr
&&- \int_0^\infty dr P_m(r) \int _r^\infty dr' \delta {\Gamma_\eta}(r') - \delta\omega,
\end{eqnarray}
where $\delta \omega=\delta (M_+ {W_\eta}_+)$ is an exact form. The second term in the expression
for $\Omega$ continues to be inconvenient, but may be cast into a more appropriate form using
the identity \cite{ku1}
\begin{widetext}
\begin{equation}
\left(\int_0^r dr' P_m(r') \times \int_r^\infty dr' \delta {\Gamma_\eta}(r')\right)' = P_m(r)
\int_r^\infty dr' \delta {\Gamma_\eta}(r') - \left(\int_0^r dr' P_m(r')\right) \delta
{\Gamma_\eta}(r).
\end{equation}
\end{widetext}
Integrating from $r=0$ to $r=\infty$, the left hand side of the above equation vanishes
identically and one finds
\begin{eqnarray}
\int_0^\infty dr P_m(r)&&\int_r^\infty dr' \delta {\Gamma_\eta}(r') =\cr\cr
&& \int_0^\infty dr \delta {\Gamma_\eta}(r)\int_0^r dr' P_m(r'),
\end{eqnarray}
so that $\Omega$ can be cast into the form
\begin{equation}
\Omega = p_+ \delta M_+ + \int_0^\infty dr P_{\Gamma_\eta} \delta {\Gamma_\eta} - \delta \omega,
\end{equation}
where we have defined
\begin{eqnarray}
p_+ = {W_\eta}_+ + \int_0^\infty dr P_m(r)\cr\cr
P_{\Gamma_\eta} = -\int_0^r dr' P_m(r').
\end{eqnarray}
Eliminating a total time derivative turns the action in (\ref{system1}) into
\begin{widetext}
\begin{equation}
S = \int dt \left(p_+ {\dot M}_+ + \int_0^\infty dr [P_{W_\eta} {\dot {W_\eta}} + \oP_R {\dot R}~ +
P_{\Gamma_\eta} {\dot {\Gamma_\eta}} - N \cH - N^r \cH_r]\right),
\end{equation}
\end{widetext}
where
\begin{eqnarray}
\cH &=& - \left[\frac{{\Gamma_\eta} \cF^{-1} R' - \cF P_{\Gamma_\eta}' \oP_R}{L} \right] - \eta \frac{P_{W_\eta}
{W_\eta}'} {L}\cr\cr
\cH_r &=& R'\oP_R - {\Gamma_\eta} P_{\Gamma_\eta}' + {W_\eta}' P_{W_\eta}.
\label{const2}
\end{eqnarray}
Furthermore, it follows from (\ref{wprime}) that
\begin{equation}
P_{\Gamma_\eta}' = {W_\eta}' + \eta \frac{R'}{\cF}.
\label{wprime2}
\end{equation}
The constraints, $\cH \approx 0 \approx \cH_r$, can be further simplified by using $\cH_r \approx 0$
to eliminate ${W_\eta}'P_{W_\eta}$ from the Hamiltonian constraint. This gives
\begin{equation}
\left(\cF P_{\Gamma_\eta}' + \eta R'\right)\left(\oP_R - \eta \frac{{\Gamma_\eta}} {\cF}\right)
\approx 0.
\end{equation}
Consider the first of the two factors above. Using (\ref{wprime2}) to substitute for $P_{\Gamma_\eta}'$,
we find
\begin{equation}
\cF P_{\Gamma_\eta}' + \eta R' = \cF \left({W_\eta}' + 2\eta \frac{R'}{\cF}\right) = \cF\frac{\oW'}
{\sigma},
\end{equation}
where $\sigma(R,{W_\eta})$ is the integrating factor introduced in the previous section. But
${\oW}' \neq 0$ because it is a null coordinate and is required to increase in time,
therefore,
\begin{equation}
\oP_R - \eta \frac{{\Gamma_\eta}}{\cF} \approx 0
\end{equation}
is equivalent to the Hamiltonian constraint, $\cH \approx 0$. Inserting this into either
of the two constraints then gives $P_{W_\eta} \approx \eta \cF \oP_R \approx {\Gamma_\eta}$.

The configuration space is made of the set of variables $\{{W_\eta},R,{\Gamma_\eta},M_+\}$, whose
physical significance is transparent. This is an advantage of Kucha\v r variables. ${W_\eta}$ is
a null coordinate, $R$ is the area radius of a point labeled $(r,t)$, ${\Gamma_\eta}$ is the energy
density of the collapsing cloud and $M_+$ is the mass measured at spatial infinity. $M_+$ is a
constant of the motion and may be viewed as part of the initial data. Our gravity-matter system
may be re-written as
\begin{widetext}
\begin{equation}
S =  \int dt \left(p_+ {\dot M}_+ + \int_0^\infty dr \left[P_{W_\eta} {\dot {W_\eta}} + \oP_R {\dot R} +
P_{\Gamma_\eta} {\dot {\Gamma_\eta}} - N^W (P_{W_\eta} - {\Gamma_\eta}) - N^r (R'\oP_R - {\Gamma_\eta}
P_{\Gamma_\eta}' + {W_\eta}' P_{W_\eta})\right]\right).
\label{system3}
\end{equation}
\end{widetext}
The canonical action (\ref{system3}) can be a starting point for Dirac quantization. The
physical meaning of all the configuration space coordinates is clear: ${W_\eta}$ and $R$ locate
the hypersurface and ${\Gamma_\eta}$ (along with $M_+$) determines the matter distribution.

\section{\label{sec:quant}Quantization}

The configuration space consists of two disconnected components, for expanding
null matter it is spanned by the set $\{W_+,R,{\Gamma_+}\,M_+\}$ and for collapsing matter by the set
$\{W_-,R,{\Gamma_-}, M_+\}$. In each case the phase-space has non-trivial boundaries,
\begin{eqnarray}
M_+ &\geq& 0,~~ P_Z \leq 0,~~ (Z=-W_\pm)\cr\cr
R &>& 0.
\label{restric}
\end{eqnarray}
The last is due to the fact that $R(t,r)=0$ describes the central singularity, {\it i.e.,} the final
singularity for the collapsing solution and the initial singularity for expanding one.
In Dirac's approach, when the phase-space admits trivial boundaries, the canonical momenta $P_X$,
are raised to operator status,
\begin{equation}
P_X \rightarrow -i \frac{\delta}{\delta X}
\label{oper}
\end{equation}
and the constraints are considered as operator restrictions on the state functional. When the
boundaries are non-trivial, as is the case in (\ref{restric}), this na\"\i ve exchange
of momenta for functional derivatives in (\ref{oper}) may lead to operators that are not
self-adjoint \cite{isham}, but we will assume here that a suitable measure on the Hilbert space
can be found, with respect to which the above replacement leads indeed to self-adjoint operators.
Subject to this caveat, the state-functional obeys
\begin{eqnarray}
-i\frac{\delta \Psi}{\delta W_\eta} &=& {\Gamma_\eta} \Psi\cr\cr
R'\frac{\delta \Psi}{\delta R} + {W_\eta}' \frac{\delta \Psi}{\delta {W_\eta}} - {\Gamma_\eta}
\left(\frac{\delta \Psi}{\delta {\Gamma_\eta}}\right)' &=& 0,
\end{eqnarray}
The last constraint is solved by any functional that is a spatial scalar. Consider a solution of
this constraint that is of the form
\begin{equation}
\Psi = C_\eta (M_+)\exp\left[i\int_0^\infty dr {\Gamma_\eta}(r) \cdot \cK (\eta,{W_\eta},R_,M_+)\right],
\label{gensol}
\end{equation}
where $C_\eta$ is a constant depending only on $\eta$ and $M_+$, and $\cK$ is an arbitrary complex
valued function of its arguments (and not their derivatives) that is to be evaluated so that
$\Psi$ satisfies the other constraints. The wave-functional $\Psi$ in (\ref{gensol}) is evidently a spatial
scalar because ${\Gamma_\eta}(r)$ is a spatial density and $\cK$ is a spatial scalar. It is therefore a
solution of the constraint providing that $\cK$ has no explicit dependence on $r$.

The solution, which agrees with all the constraints considered as operator restrictions on
the state functional, is given by
\begin{equation}
\Psi = C_\eta \exp\left[i\int_0^\infty dr {\Gamma_\eta}(r) ({W_\eta}(r) + \eta R_*(r))\right],
\end{equation}
where $R_*(R,M_+,{\Gamma_\eta})$ is a ``tortoise''-like coordinate defined by
\begin{equation}
R_*=R+2m\ln |\frac{R}{2m}-1|.
\label{rstar}
\end{equation}
It is not, of course, the tortoise coordinate $R^*$ except in a Schwarzschild region when
$m(M_+,{\Gamma_\eta})=M_+ $ is constant.

The parameter $\eta$ represents the direction of the flow, being $+1$ for outgoing null
matter and $-1$ for incoming matter. The combination $\eta R(r)$ represents the affine
parameter, $\xi(r)$. Re-expressing the wave-functional in (\ref{gensol}) to make
the dependence on the affine parameter explicit, we find
\begin{equation}
\Psi = C_\eta e^{i\int_0^\infty dr {\Gamma_\eta}(r)\xi_*(r)} \exp\left[i\int_0^\infty dr
{\Gamma_\eta}(r)\cdot {W_\eta}(r)\right],
\end{equation}
where $\xi_*=\eta R_*$.

For $\eta=-1$, $\xi \in (-\infty,0)$ and ${W_\eta}=V$, the functionals are of the form
\begin{equation}
\Psi_{-1} = A(M_+)e^{-i\int_0^\infty dr {\Gamma_-}(r)R_*(r)} e^{i\int_0^\infty dr {\Gamma_-}(r) V(r)},
\label{minus}
\end{equation}
and describe collapsing null matter. Likewise, for $\eta=+1$, $\xi \in (0,\infty)$ and ${W_\eta}=U$,
the functionals
\begin{equation}
\Psi_{+1} = B(M_+) e^{i\int_0^\infty dr {\Gamma_+}(r)R_*(r)} e^{i\int_0^\infty dr {\Gamma_+}(r)U(r)},
\label{plus}
\end{equation}
describe expanding null matter.

A given classical collapse problem is specified by a choice of mass function, $m({W_\eta})$, which
determines an initial energy distribution, thus a collapse ``model''. As an example, we shall consider
the function,
\begin{equation}
m({W_\eta}) = M_+ \theta ({W_\eta}-w),
\end{equation}
where $\theta$ is the Heaviside unit step-function and $w$ is constant. The matter energy vanishes when
${W_\eta}<w$ and is $M_+$ when ${W_\eta}\geq w$. The mass function evidently makes sense only as a thin
shell that is collapsing toward the center and we must have ${W_\eta}=V$ ($\eta=-1$). Calling the
corresponding mass at spatial infinity $M^{\rm in}_+$, we find the energy density by differentiating
w.r.t. the ADM label coordinate, $r$,
\begin{eqnarray}
\Gin(r) &=& M^{\rm in}_+ V'(r)\delta(V-v)\cr\cr
&=& M^{\rm in}_+ \frac{V'(r)}{V'(\zr)}\delta(r-\zr),
\end{eqnarray}
where $\zr(t)$ is the solution of $V(r,t)=v$. Likewise, a thin shell that expands out of the center
is represented by the mass function
\begin{equation}
m({W_\eta}) = M^{\rm out}_+ \theta(w-{W_\eta}),
\end{equation}
for ${W_\eta}=U$ ($\eta=+1$). The energy density is
\begin{eqnarray}
\Gout(r)&=&-M^{\rm out}_+ U'(r)\delta(u-U)\cr\cr
&=& M^{\rm out}_+ \frac{U'(r)}{U'(\zr)}\delta(r-\zr),
\end{eqnarray}
where $\zr(t)$ is the solution of $U(r,t)=u$. The energy density is always positive and in either
case we find
\begin{equation}
\Gamma=M^{\rm in(out)}_+\delta(r-\zr).
\label{shelldensity}
\end{equation}
The shell trajectories, $\zr = \zr(t)$ are different in the two cases, being $v={\rm const.}$ in the
first and $u={\rm const.}$ in the second. Using these expressions, let us relate the constraints in
(\ref{system}) to the constraints that have been used by others to describe thin shells. The dust
Hamiltonian and momentum density turn out to be ($P_Z=-\Gamma$)
\begin{eqnarray}
\cH^d &=& \eta \frac{M_+ Z'}{L} \delta(r-\zr) = \eta \frac{\zp}{L} \delta(r-\zr)\cr\cr
\cH^d_r &=& -M_+ Z' \delta(r-\zr) = -\zp \delta(r-\zr),
\end{eqnarray}
where we have defined $\zp = M_+ Z'(\zr)$ and therefore $\eta={\rm sgn}(Z')={\rm sgn}(\zp)$. These
expressions were used as a starting point in \cite{Louko,haj1,haj2}. The gravitational contributions to
the constraints are, of course, the same.

The corresponding classical solutions are represented in the Penrose diagram of figure 2, where
$AA'$ represents the event horizon in (a) and the Cauchy horizon in (b).
\begin{figure}
\epsfysize=5cm \epsfbox{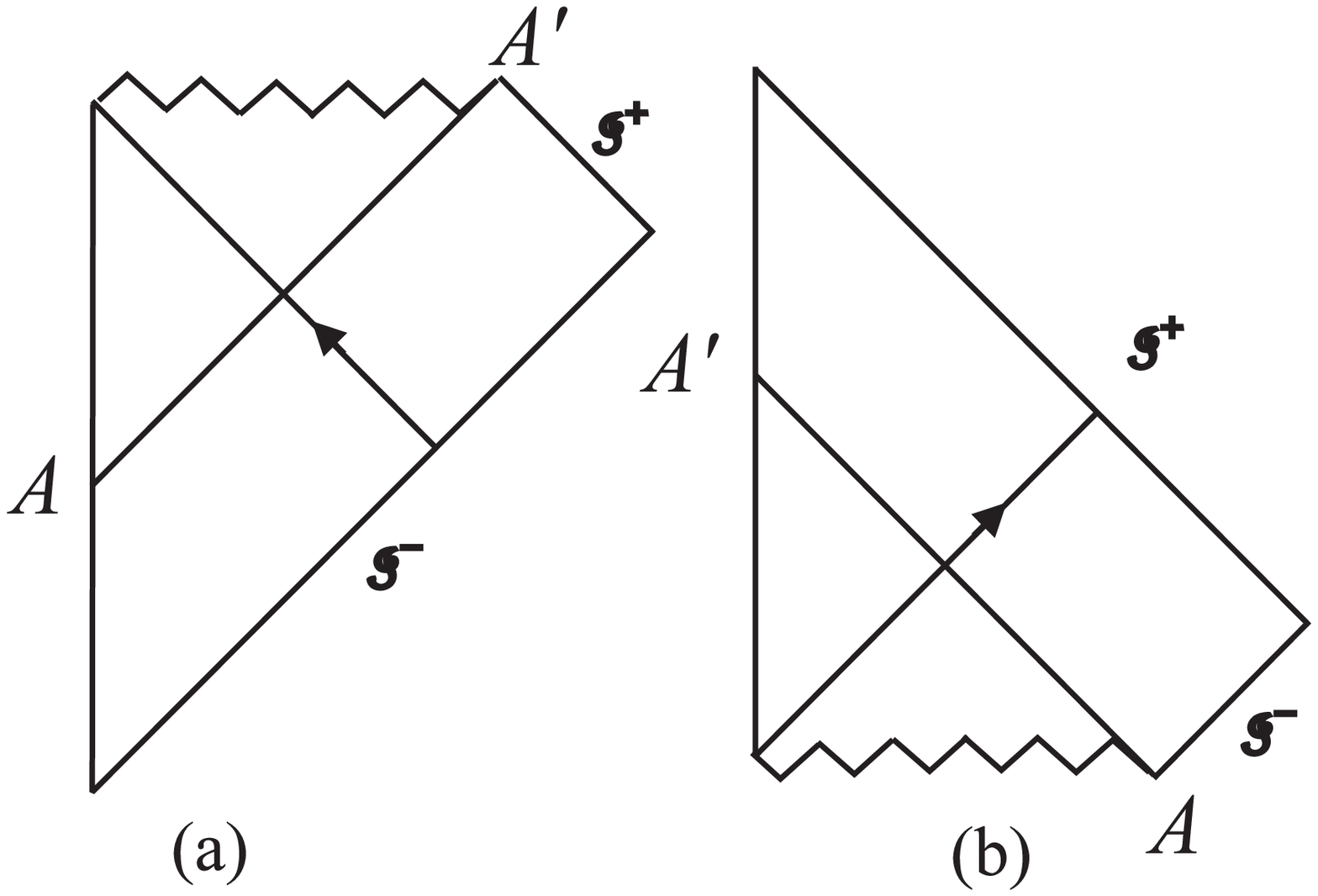}
\caption{A thin shell (a) collapsing and (b) expanding}
\end{figure}
Inserting (\ref{shelldensity}) into (\ref{minus}) and (\ref{plus}) one finds that the
quantum mechanics of a single collapsing shell is defined on the reduced configuration space
$(v,R,M^{\rm in}_+)$ and described by the wave-function
\begin{equation}
\Psi_{-1} = A(M_+^{\rm in},v)e^{-iM_+^{\rm in}R^*(\zr)},
\end{equation}
whereas, for an outgoing shell the reduced configuration space is $(u,R,M^{\rm out}_+)$ and the
wave-function
\begin{equation}
\Psi_{+1} = B(M_+^{\rm out},u)e^{iM_+^{\rm out}R^*(\zr)},
\end{equation}
where $R^*$ as given by (\ref{rstar}) is, in this case, the usual tortoise coordinate.

\section{\label{sec:disc}Discussion}

In this paper we have examined the collapse and expansion of a null dust cloud of arbitrary mass
distribution and shown that there exists a canonical transformation that brings the corresponding
Vaidya system to the Kucha\v r form, in which the dynamics is expressed in terms of embedding
variables whose physical meaning is transparent. Written in terms of these variables, the
constraints take on a simple form.

The physical content of the solution wave-functionals discussed in the previous section is given
by the expectation values of observables. These are, for example, the geometric invariants and matter
invariants (such as curvature scalars and trace of the stress energy tensor) and in general any
$C^{\infty}$ function of the phase-space variables, old or new, that weakly commutes with the
constraints.  They must be written as operators on the Hilbert space, but there are two associated
difficulties. Firstly, because they are generally non-linear, there will be operator ordering
ambiguities. Secondly, one must ensure that they are self-adjoint with respect to the chosen measure.
However, as our transformation between the spaces is explicit, knowing these functions in one
system is equivalent to knowing them in the other. It should be noted that a complete set of
Dirac observables has been constructed in the case of a single shell \cite{haj1} where a single
degree of freedom is present. For the general case, this issue is quite complicated and will be
discussed in the future.

Subject to the condition that suitable boundary conditions or a suitable inner product can be
found so that (\ref{oper}) leads to self-adjoint operators, we have found solution wave-functionals
for arbitrary mass distributions in each sector (collapse and expansion) independently (as mentioned
in the introduction, solutions with non-trivial mass distributions are essential for the description
of issues as naked singularities and the Cosmic Censor). These sectors are disjoint,
separated by the central singularity (at $R=0$). However, if the solutions are viewed as describing the
evolution, in the affine parameter $\xi$, of a collapsing matter cloud beginning on ${\frak I}^{-}$,
they seem to suggest that it should be possible to define the wave-functional over the entire
interval $\xi \in (-\infty, \infty)$ by extending the range of $R$ to include the center.
Yet, for any matter distribution, the solution space-time in (\ref{sptime}) admits a
strong curvature singularity at the origin, so the classical dynamics cannot be extended even to
it. Any attempt to continue the quantum dynamics through the origin must therefore ensure
that at least the expectation values of the observables in the consequent quantum theory are well
behaved there. Thus the central singularity would be made harmless by the quantum theory. It can
be so if, for example, the wave-functional were to vanish there. In this way the matter would collapse
and re-expand through the (benign) center in one continuous history, the solutions being given by
(\ref{minus}) for $\xi \in (-\infty,0]$ and (\ref{plus}) for $\xi \in [0,\infty)$. This has been
proposed for a single shell by H\'aj\'\i \v cek and Kiefer \cite{haj1,haj1a,haj2}, who merged the
two solutions into one bouncing solution. Their bounce was obtained by working in double-null
coordinates and employing group quantization techniques (see \cite {haj1} and \cite{haj3}) to the
problem. Group quantization is beautifully adapted to the quantization of systems with non-trivial
boundary restrictions such as those in (\ref{restric}) on the phase space, but it's application
to problems with more than a few number of degrees of freedom and in particular to the collapse of
general matter distributions, being dependent on the construction of a complete set of observables,
remains a topic for future investigation.

The Eddington-Finkelstein coordinates we have employed in this paper present several advantages over
the double null system in regard to the problem of collapse or re-expansion of arbitrary matter
distributions. The new variables have a clear physical and geometrical meaning. This is useful
when comparing the quantum behavior with the classical. Our transformations are explicit, which
means that operators and, in particular, observables that are known in one coordinate system can be
expressed in the other. The matter-gravity constraints in the new phase space are linear for all
matter distributions. This simplification, achieved on the classical level, has allowed us to obtain
exact solution wave-functionals describing the respective physical processes.

\section{\label{sec:ack}Acknowledgements}

\noindent We acknowledge the partial support of FCT, Portugal, under contract number POCTI/32694/FIS/2000.
L.W. was supported in part by the Department of Energy, USA, under contract Number DOE-FG02-84ER40153.



\begin{thebibliography}{99}
\bibitem{hawk1}S. W. Hawking, Commun. Math. Phys. {\bf 43} (1975) 199.
\bibitem{bek1}J. D. Bekenstein, Phys. Lett. {\bf B481} (2000) 339; J. D. Bekenstein and A. Mayo, Phys,
Phys. Rev. {\bf D61} (2000) 024022; S. Hod, Phys. Rev. {\bf D61}, 024023 (2000); J. D. Bekenstein and M. Schiffer,
Phys. Rev. {\bf D58} (1998) 064014; J. D. Bekenstein, Phys. Lett. {\bf B360} (1995) 7; {\it ibid} Phys.
Rev. Lett. {\bf 70} (1993) 3680; {\it ibid} Phys. Rev. {\bf D9} (1974); {\it ibid} Phys. Rev. {\bf D7}
(1973) 2333; {\it ibid}, Ph. D. Thesis, Princeton University, April 1972; {\it ibid}, Lett. Nuovo Cimento
{\bf 4} (1972) 737.
\bibitem{av}K. V. Krasnov, Gen. Rel. Grav. 30 (1998) 53; A. Ashtekar, J. Baez, A. Corichi, K Krasnov,
Phys. Rev. Letts. {\bf 80} (1998) 904; A. Ashtekar, J.Lewandowski, Class. Quant. Grav. {\bf 14} (1997)
A55; H. A. Kastrup, Phys. Lett. {\bf B385} (1996) 75; {\it ibid} Phys. Letts. {\bf B413} (1997) 267;
{\it ibid} Phys. Letts. {\bf B419} (1998) 40.
\bibitem{st}A. W. Peet, Class. Quant. Grav. {\bf 15} (1998) 3291; K. Sfetsos, K. Skenderis, Nucl. Phys.
{\bf B517} (1998) 179; A. Strominger and C. Vafa, Phys. Lett. {\bf B379} (1996) 99;  C. O. Lousto, Phys.
Rev. {\bf D51} (1995) 1733; M. Maggiore, Nucl. Phys. {\bf B429}  (1994) 205;  Ya. I. Kogan, JETP Lett.
{\bf 44} (1986) 267; A. Strominger, JHEP 9802 (1998) 009; J.M. Maldacena and A. Strominger, JHEP 9802
(1998) 014; J.M. Maldacena, A. Strominger and E. Witten, JHEP 9712 (1997) 002; G. Horowitz, D. Lowe and
J.M. Maldacena, Phys. Rev. Lett. {\bf 77} (1996) 430-433.
\bibitem{cqg}J. M\"akel\"a, P. Repo, M. Luomajoki and J. Piilonen, Phys. Rev. {\bf D64} (2001) 024018; Cenalo
Vaz and Louis Witten, Phys. Rev. {\bf D64} (2001) 084005; C. Kiefer and J. Louko, Annalen Phys. 8 (1999) 67;
C. Kiefer, Nucl. Phys. {\bf B} Proc. Suppl 57 (1997) 173; T. Brotz and C. Kiefer, Phys. Rev. {\bf D55} (1997)
2186; J. Louko and B.F. Whiting, Phys. Rev. {\bf D51} (1995) 5583; J. Louko, S. Winters-Hilt, Phys. Rev.
{\bf D54} (1996) 2647; F. Larsen and F. Wilczek, Phys. Letts. {\bf B375} (1996) 37; J. Louko and J. M\"akel\"a,
Phys. Rev. {\bf D54} (1996) 4982; P. Kraus and F. Wilczek, Nucl. Phys. {\bf B437} (1995) 231; F. Larsen and
F. Wilczek, Annals Phys. 243 (1995) 280.
\bibitem{Rp1}R. Penrose, Riv. Nuovo Cimento {\bf 1} (1969) 252; in {\it General Relativity, An Einstein
Centenary Survey}, ed. S. W. Hawking and W. Israel, Cambridge Univ. Press, Cambridge, London (1979)
581.
\bibitem{Pj1}see, for example, P. S. Joshi, {\it Global Aspects in Gravitation and
Cosmology}, Clarendon Press, Oxford, (1993).
\bibitem{cvetc}T.P. Singh and Cenalo Vaz, Phys. Rev. {\bf D61} (2000) 124005; T.P. Singh and Cenalo
Vaz, Phys. Letts. {\bf B481} (2000) 74; S. Barve, T.P. Singh, Cenalo Vaz and Louis Witten, Nucl. Phys.
{\bf B532} (1998) 361; S. Barve, T.P. Singh, Cenalo Vaz and Louis Witten, Phys. Rev. {\bf D58}
(1998) 104018; S. Barve, T.P. Singh and Cenalo Vaz, Phys. Rev. {\bf D62} (2000) 084021; T. Harada,
H. Iguchi and K. Nakao, Phys. Rev. {\bf D61}, 101502 (2000); T. Harada, H. Iguchi and K. Nakao, Phys.
Rev. {\bf D62}, 084037 (2000); Cenalo Vaz and Louis Witten, Phys. Lett. {\bf B442} (1998) 90.
\bibitem{haretc}T. Harada, H. Iguchi, K. Nakao, T.P. Singh, T. Tanaka and Cenalo Vaz, Phys. Rev.
{\bf D64} (2001) 041501.
\bibitem{vaidya}P.C. Vaidya, Proc. Indian. Acad. Sci. {\bf A33}, 264.
\bibitem{nsing} W. A. Hiscock, L. G. Williams and D. M. Eardley (1982) Phys. Rev. {\bf D26} 751; Y.
Kuroda (1984) Prog. Theor. Phys. {\bf 72} 63; A. Papapetrou (1985) in A Random Walk in General Relativity,
Eds. N. Dadhich, J. K. Rao, J. V. Narlikar and C. V. Vishveshwara (Wiley Eastern, New Delhi); G. P.
Hollier (1986) Class. Quantum Grav. {\bf 3} L111; W. Israel (1986) Can. Jour. Phys. {\bf 64} 120; K.
Rajagopal and K. Lake (1987) Phys. Rev. {\bf D35} 1531; I. H Dwivedi and P. S. Joshi (1989) Class.
Quantum Grav. {\bf 6} 1599; (1991) Class. Quantum Grav. {\bf 8} 1339; P. S. Joshi and I. H. Dwivedi
(1992) Gen. Rel. Gravn. {\bf24} 129; J. Lemos (1992) Phys. Rev. Lett. {\bf 68} 1447.
\bibitem{ku1}K. V. Kucha\v r, Phys. Rev. {\bf D50} (1994) 3961.
\bibitem{isham}C.J. Isham in ``Relativity, groups and topology II'', ed. B.S. deWitt and R. Stora,
Elsevier, Amsterdam, (1984).
\bibitem{Louko}J. Louko, B. Whiting and J. Friedman, Phys. Rev. {\bf D57} (1998) 2279.
\bibitem{kuBi}J. Bi\v c\'ak and K.V. Kucha\v r, Phys.Rev. {\bf D56} (1997) 4878.
\bibitem{haj1}P. H\'aj\'\i \v cek, Nucl. Phys. {\bf B603} (2001) 555-577.
\bibitem{haj1a}P. H\'aj\'\i \v cek and C. Kiefer, gr-qc/0107102.
\bibitem{haj2}P. H\'aj\'\i \v cek and C. Kiefer, Nucl. Phys. {\bf B603} (2001) 531-554.
\bibitem{haj3}P. H\'aj\'\i \v cek, J. Math. Phys. {\bf 36} (1995) 4612; P. H\'aj\'\i \v cek,
A Iguchi and J. Tolar, J. Math. Phys. {\bf 36} (1995) 4639.
\end{thebibliography}
\end{document}